\documentclass[namedreferences]{SolarPhysics}
\usepackage[optionalrh]{spr-sola-addons} % For Solar Physics 
\usepackage{graphicx}        % For eps figures, newer & more powerfull
\usepackage{color}           % For color text: \color command
\usepackage{url}             % For breaking URLs easily trough lines
\begin{document}

\begin{article}

\begin{opening}

\title{An Automated Algorithm to Distinguish and Characterize Solar Flares and Associated Sequential Chromospheric Brightenings}
\author{M.~S.~\surname{Kirk}$^{1,2,3}$\sep
        K.~S.~\surname{Balasubramaniam}$^{2,3,1}$\sep
        J.~\surname{Jackiewicz}$^{1}$\sep
        B.~J.~\surname{McNamara}$^{1}$\sep
        R.~T.~J.~\surname{McAteer}$^{1}$      
       }
\runningauthor{Kirk et al.}
\runningtitle{Tracking SCBs and Associated Flare Ribbons}

   \institute{$^{1}$ Department of Astronomy, New Mexico State University, Las Cruces, NM 88003\\
                     email:~\url{mskirk@nmsu.edu} email:~\url{jasonj@nmsu.edu} email:~\url{bmcnamar@nmsu.edu} email:~\url{mcateer@nmsu.edu} \\ 
              $^{2}$ Space Vehicles Directorate, Air Force Research Laboratory, Sunspot, NM 88349\\
              $^{3}$ National Solar Observatory, Sunspot, NM 88349\\
             }

\begin{abstract}
We present a new automated algorithm to identify, track, and characterize small-scale brightening associated with solar eruptive phenomena observed in H$\alpha$. The temporal spatially-localized changes in chromospheric intensities can be separated into two categories: flare ribbons and sequential chromospheric brightenings (SCBs). Within each category of brightening we determine the smallest resolvable locus of pixels, a kernel, and track the temporal evolution of the position and intensity of each kernel. This tracking is accomplished by isolating the eruptive features, identifying kernels, and linking detections between frames into trajectories of kernels. We fully characterize the evolving intensity and morphology of the flare ribbons by observing the tracked flare kernels in aggregate. With the location of SCB and flare kernels identified, they can easily be overlaid on top of complementary data sets to extract Doppler velocities and magnetic field intensities underlying the kernels. This algorithm is adaptable to any dataset to identify and track solar features. 
\end{abstract}
\keywords{Flares, Dynamics; Chromosphere, Active}
\end{opening}
%-------------------------------------------------
\section{Introduction}
     \label{S-Introduction} 
     
Active regions on the Sun can produce a variety of distinct dynamic features including sunspots, flares, prominences and occasionally mass ejections; as such solar active regions are a major source of adverse space weather.  Historically, these active regions have been identified and characterized manually. To better understand the dynamics of solar active regions and their associated phenomena, data from different sources must be utilized and a large dataset consisting of numerous active regions must be analyzed. This necessitates an automated approach to correlate several seemingly disparate data sets. 

In the past, a few different automated methods have been developed to simultaneously analyze flare ribbons across multiple sources of data \cite{Qu2003,Maurya2010,Gill2010}, however sequential chromospheric brightenings (SCB) have not been studied in such an automated fashion.  First observed in 2005, sequential chromospheric brightenings (SCBs) appear as a series of spatially separated points that brighten in sequence \cite{Bala2005}. The sequential nature of the point brightening gives the appearance of a progressive traveling disturbance. SCBs are observed as both single and multiple trains of brightenings in association with a large-scale eruption in the chromosphere or corona. The loci of brightenings are seen in a time-series of full-disk H$\alpha$ images as emerging predominantly along the axis of the flare ribbons. Physically, SCBs are closely correlated with solar flares, coronal restructuring of magnetic fields, halo CMEs, EIT waves, and chromospheric sympathetic flaring \cite{Bala2005}. \inlinecite{Pevtsov2007} show that these physical correlations are consistent with several properties of chromospheric evaporation.

This paper presents a new automated method of identifying and tracking SCBs and associated flare ribbons. This tracking technique is different than previous flare tracking algorithms in that it identifies and tracks subsections of the flare from pre-flare through the impulsive brightening and the exponential decline. This allows measurement of the temporal variation in intensity and position, which can then be combined with other data to infer the changes occurring to the underlying photospheric magnetic field as well as derived H$\alpha$ Doppler velocity of each individual flare kernel. This tracking algorithm is also adapted to identify and track the temporal evolution of the ephemeral SCBs associated with the flaring region. Section~\ref{S-Data} describes the data sets used to train the algorithm and the image processing involved before the detection routine can be applied. Section~\ref{S-Analysis} explains the method of detection used to identify and track the flare and SCB kernels. Section~\ref{S-Results} presents the capabilities of applying this detection algorithm to flaring events. Finally, Section~\ref{S-Discuss} discusses the implications of these results and the future direction of this project. 

\section{Data}
     \label{S-Data} 
\begin{figure}    %%%%%%%%%%%%%%%%%% FIGURE 1 
   \centerline{\includegraphics[width=1.3\textwidth,clip=,angle=90]{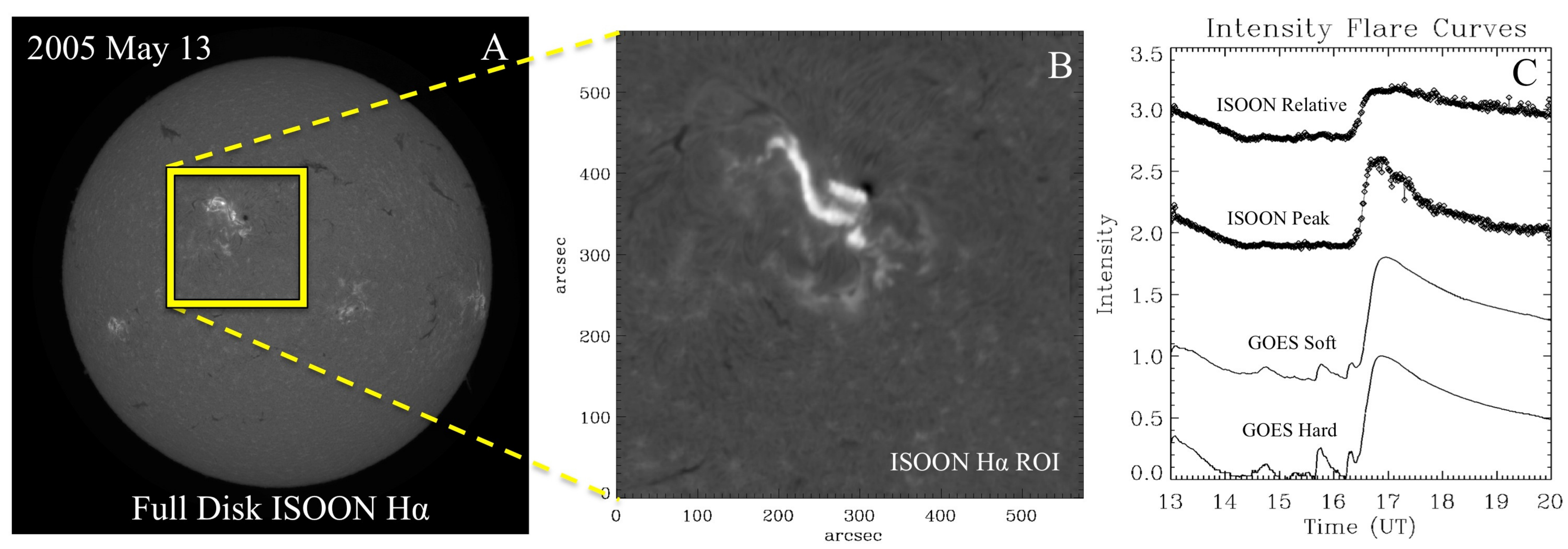}
              }
              \caption{The two-ribbon flaring event from May 13, 2005. {\bf A} shows an example of a calibrated H$\alpha$ ISOON image with the region of interest (ROI) highlighted. The ISOON images are 2048 x 2048 pixels with each pixel having a 1.1 arcsec resolution. {\bf B} is the ROI after preprocessing. This is ROI covers 576 x 576 pixels corresponding to $2.12 \times 10^{11}$ km$^2$ on the solar surface.  {\bf C} shows intensity curves over the time period of interest: 13:00 -- 20:02 UT.  The ISOON Relative curve is an intensity curve defined by Equation~\ref{Eq-Contrast} while the ISOON Peak curve is defined by the maximum value of the ROI in each time step; both curves are normalized. The GOES Soft ($1.0-8.0$ \AA) and Hard ($0.5 - 4.0$ \AA) curves are normalized x-ray intensity curves for reference from the GOES 14 satellite.
                      }
   \label{0513_boxes}
   \end{figure}

\begin{figure}    %%%%%%%%%%%%%%%%%% FIGURE 2 
   \centerline{\includegraphics[width=1.3\textwidth,clip=,angle=90]{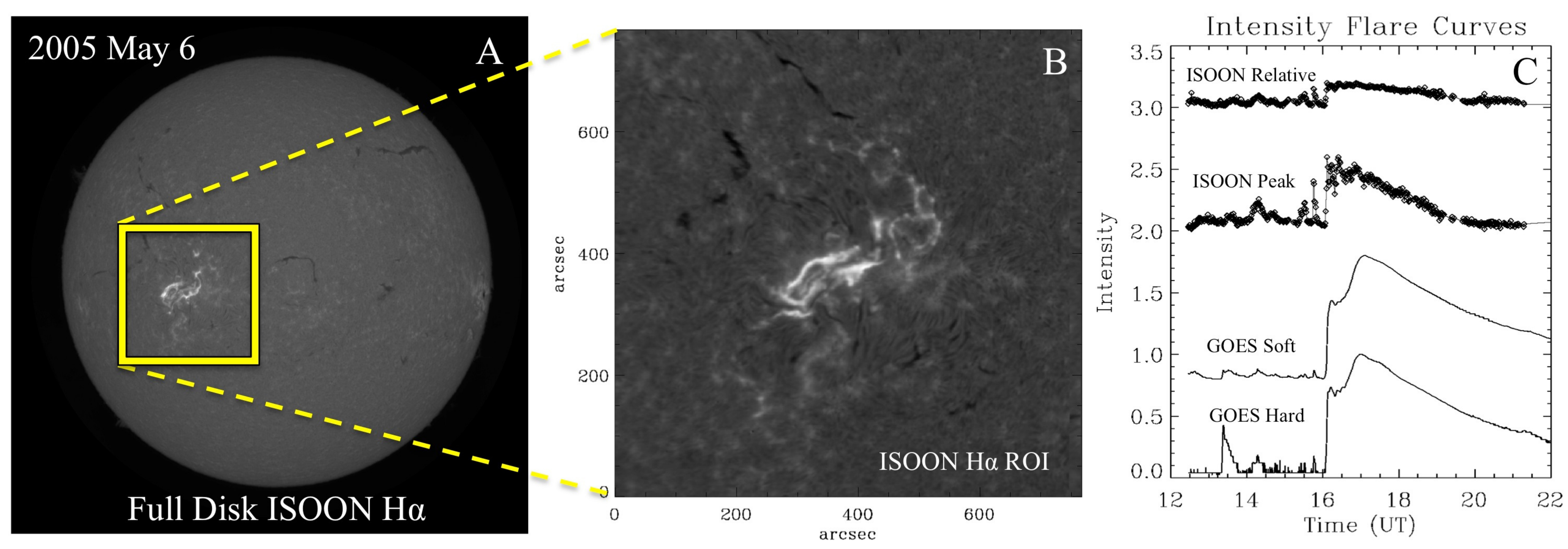}
              }
              \caption{The two-ribbon flaring event from May 06, 2005. {\bf A} shows an example of a calibrated H$\alpha$ ISOON image with the region of interest (ROI) highlighted. The ISOON images are 2048 x 2048 pixels with each pixel having a 1.1 arcsec resolution. {\bf B} is the ROI after preprocessing. This is ROI covers 768 x 768 pixels corresponding to $3.77 \times 10^{11}$ km$^2$ on the solar surface.  {\bf C} shows intensity curves over the time period of interest: 12:28 -- 22:47 UT.  The ISOON Relative curve is an intensity curve defined by Equation~\ref{Eq-Contrast} while the ISOON Peak curve is defined by the maximum value of the ROI in each time step; both curves are normalized. The GOES Soft ($1.0-8.0$ \AA) and Hard ($0.5 - 4.0$ \AA) curves are normalized x-ray intensity curves for reference from the GOES 14 satellite.
                      }
   \label{0506_boxes}
   \end{figure}

\begin{figure}    %%%%%%%%%%%%%%%%%% FIGURE 3 
   \centerline{\includegraphics[width=1.2\textwidth,clip=,angle=90]{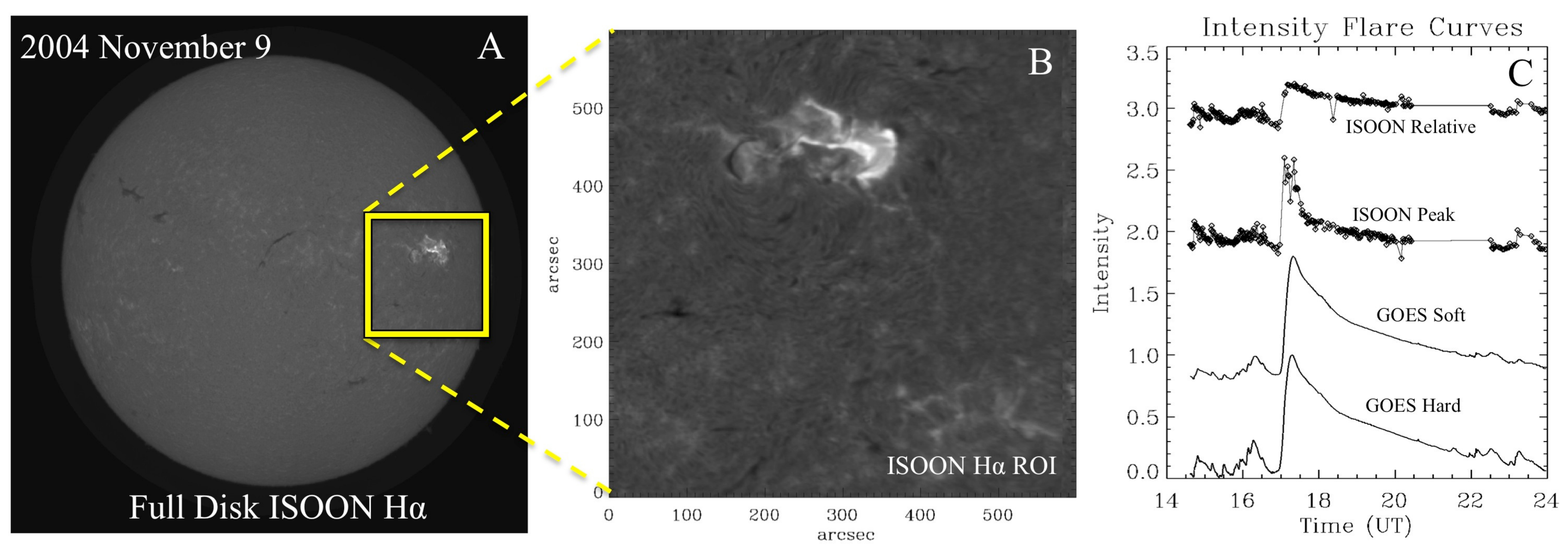}
              }
              \caption{The two-ribbon flaring event from November 9,2004. {\bf A} shows an example of a calibrated H$\alpha$ ISOON image with the region of interest (ROI) highlighted. The ISOON images are 2048 x 2048 pixels with each pixel having a 1.1 arcsec resolution. {\bf B} is the ROI after preprocessing. This is ROI covers 600 x 600 pixels corresponding to $2.30 \times 10^{11}$ km$^2$ on the solar surface.  {\bf C} shows intensity curves over the time period of interest: 14:37 -- 23:58 UT.  The ISOON Relative curve is an intensity curve defined by Equation~\ref{Eq-Contrast} while the ISOON Peak curve is defined by the maximum value of the ROI in each time step; both curves are normalized. The GOES Soft ($1.0-8.0$) and Hard ($0.5 - 4.0$ \AA) curves are normalized x-ray intensity curves for reference from the GOES 14 satellite.
                      }
   \label{1109_boxes}
   \end{figure}

This project utilizes H$\alpha$  (6563 \AA) images from the Improved Solar Observing Optical Network (ISOON) telescope to train the algorithm in identifying and tracking flare ribbons and SCBs. ISOON produces a 2048 by 2048 pixel full-disk image with a one-minute cadence \cite{Neidig1998}. These full-disk chromospheric images have a 1.1 arc-second resolution and are normalized for quiet Sun intensity and corrected for atmospheric refraction (Figures~\ref{0513_boxes}{\bf A}, \ref{0506_boxes}{\bf A}, and \ref{1109_boxes}{\bf A}). At the same resolution and cadence, $\pm0.4$ \AA\ off-band Doppler images are also recorded. 

In this study, we chose three flares (Figures~\ref{0513_boxes}, \ref{0506_boxes}, and \ref{1109_boxes}) with previously identified SCB events by \inlinecite{Bala2006}  as a sample set to train the detection algorithm (see Table~\ref{T-events}). Each of the events selected had halo CMEs associated with the flare, a filament eruption, and had a rough two-ribbon topology. Both 6 and 13 May 2005 events were located near disk center while the 9 November 2004 event was near the eastern limb. To achieve complete coverage of the flare's eruption, images from $\pm 3.5$ hours from each eruption's start time are extracted from the archive. This yields an image cube of $\sim$400 images for an individual event. To characterize the flare eruption, three methods were used: a relative intensity curve, a maximum intensity curve, and the Geostationary Operational Environmental Satellite (GOES) hard and soft x-ray fluxes (Figures~\ref{0513_boxes}{\bf C}, \ref{0506_boxes}{\bf C}, and \ref{1109_boxes}{\bf C}). The relative intensity curve is a ratio of the maximum to mean intensity for each of the H$\alpha$ images:
\begin{equation}
	\label{Eq-Contrast}
I_r=\frac{I_{\rm max}-I_{\rm mean}}{I_{\rm max}+I_{\rm mean}},
\end{equation}
where $I_r$ is the relative intensity, $I_{\rm max}$ is the maximum intensity, and $I_{\rm mean}$ is the mean intensity of each H$\alpha$ image \cite{McAteer2002,McAteer2003}. The maximum intensity curve is $I_{\rm max}$ for each image. The GOES Soft ($1.0-8.0$ \AA) and Hard ($0.5-4.0$ \AA) curves are x-ray intensity curves from the GOES 14 satellite. 

\begin{table}
\caption{ 
The events selected by this study to analyze. Each event was previously identified as having SCBs associated with the flare and had a two-ribbon topology. 
}
\label{T-events}

\begin{tabular}{lccc}     % define the column alignment
  \hline                   % horizontal line
 Date & Start Time (UT) & Flare Class & CME \\
  \hline
2004 November 9& 16:30  & M & Halo \\
2005 May 5& 16:03 &  C & Halo \\
2005 May 13& 16:33 &  M & Halo \\

  \hline
\end{tabular}
\end{table}

\begin{figure}    %%%%%%%%%%%%%%%%%% FIGURE 4
   \centerline{\includegraphics[width=1.3\textwidth,clip=,angle=90]{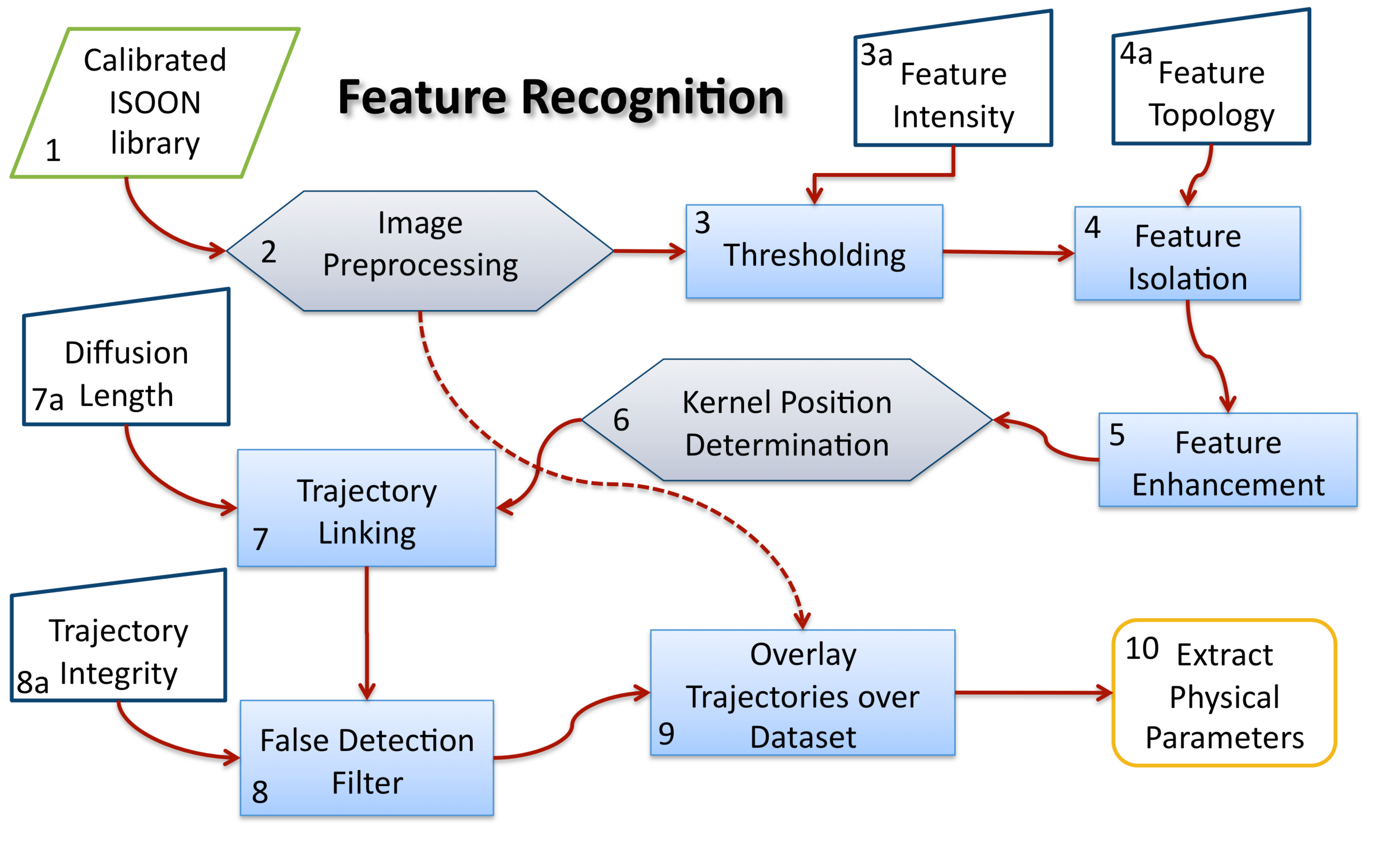}
              }
              \caption{A flowchart detailing the process in which flare ribbons and SCBs are identified and their physical parameters measured. The green parallelogram represents the database input. Blue trapezoids indicate a feature specific quality that must be determined and input. The gray diamonds are processes that have been defined by a collection of subroutines that are described in this paper and the blue shaded rectangles are individual routines. 
                      }
   \label{flow_chart}
   \end{figure}

Before the detection analysis is applied, each image is preprocessed (see Figure~\ref{flow_chart}). Preprocessing of line-center H$\alpha$ images begins by removing limb darkening from the calibrated full disk ISOON image. This is accomplished by subtracting a standard limb darkening profile determined by the spherical geometry of the solar surface \cite{Hestroffer1998,Nickel1994}.  Each image is then de-projected into conformal cylindrical coordinates which removes the projection effects of imaging the solar sphere. As a consequence, latitudes above 85 degrees north or south are significantly distorted and are cropped off. In this cylindrical projection, the entire solar limb is distorted. Any region of interest at the limb is significantly distorted as well leading to added noise in the analysis; for an example of an event on the limb, see Figure~\ref{1109_boxes}.  The de-projected image is then normalized to the mean background level of the quiet Sun so that any flare brightening is comparable between images.  Next, the full-disk de-projected image is cropped down to isolate the flaring region (Figures~\ref{0513_boxes}{\bf B}, \ref{0506_boxes}{\bf B}, and \ref{1109_boxes}{\bf B}).  The flaring regions in each image are aligned using a cross-correlation algorithm. This eliminates the rotation effects of the Sun so the flare can be observed as if it was stationary on the solar surface. An algorithm to de-stretch each image is applied to the image cube to co-align subsections of the image. The mean displacement of any given subsection due to de-stretching over time is nearly zero. A net zero displacement implies the algorithm is not removing any proper motion observed in the time series. A discussion of the methodology behind the de-stretching algorithm can be found in~\inlinecite{November1988}. The entire image preprocessing of a flare is automated up to this point, given a library of ISOON images and the coordinates of the active region that is of interest. The last step in preprocessing is removing bad frames. Images that contain peculiar observations due to clouds or other factors are discarded interactively.

\section{Tracking Algorithm}
     \label{S-Analysis} 
When viewing a sequential time series of images covering an erupting flare, several physical characteristics of the evolving ribbons are immediately apparent: the ribbons separate, brighten, and change their morphology. Along with the erupting flare, SCBs can be observed brightening and dimming in the vicinity of the ribbons.  This section describes the techniques and methods used to extract quantities of interest such as location, velocity, and intensity of the flare ribbons and SCBs.  Section~\ref{S-Detect} describes the technique used to identify and track bright kernels in a set of images. In this context, we are defining a kernel to be a small locus of pixels that are associated with each other through increased intensity as compared with the immediately surrounding pixels.  Each kernel has a local maximum, must be separated from another kernel by at least one pixel, and does not have any predetermined size or shape. Section~\ref{S-Flare_anal} describes some of the image processing required to extract characteristics of flare kernels and Section~\ref{S-SCB_anal} describes the technique for SCB kernels. 

\subsection{General Kernel Detection and Tracking}
	\label{S-Detect}
The algorithm developed to identify, track and extract physical quantities in the ISOON dataset does so in seven logical steps: thresholding, feature isolation, feature enhancement, locating the position of candidate kernels,  eliminating ``false" kernels, linking time-resolved kernels together into trajectories, and extracting physical quantities by overlaying the trajectories over other datasets (numbers 3 - 9 in Figure~\ref{flow_chart}). To aid in this detection, tracking software developed by Crocker, Grier and Weeks \cite{Crocker1996} was used as a foundation and modified to fit the needs of this project. Crocker's software is available online at \url{www.physics.emory.edu/~weeks/idl/}.  

Analysis begins with thresholding the preprocessed H$\alpha$ images. A threshold is taken of the background-normalized images eliminating any features that are dimmer than a specified intensity.  Because the mean intensity of the flare ribbons are significantly brighter than SCBs, the specific thresholding level is different for each feature extraction. Section~\ref{S-Flare_anal} describes the process for flares and Section~\ref{S-SCB_anal} for SCBs.

Next, to isolate features a spatial bandpass filter is applied to suppress pixel noise with a characteristic length scale of a pixel. Noise reduction is necessary to remove the smallest changes in brightening while preserving the general structure of the images.  Selecting the length scale for the bandpass filter requires prior knowledge of the features' characteristics. With ISOON images, flare ribbons appear as $\sim$10 pixels wide and $\sim$100 pixels long - significantly larger than the bandpass filter. SCBs appear characteristically smaller in ISOON images with diameters of a few ($3-8$) pixels. The noise filter eliminates some of the smallest SCBs but is a worthwhile tradeoff for reduced noise levels. 

The process of feature enhancement is different whether looking at SCBs or flare ribbons. In either case, it is a process of folding a processed image back into the original to enhance either the flare ribbon or the SCB. Section~\ref{S-Flare_anal} describes the process for flare tracking and Section~\ref{S-SCB_anal} for SCBs.    

\subsubsection{General Kernel Position Determination}
	\label{kernel_position}
\begin{figure}    %%%%%%%%%%%%%%%%%% FIGURE 5
   \centerline{\includegraphics[width=1.3\textwidth,clip=,angle=90]{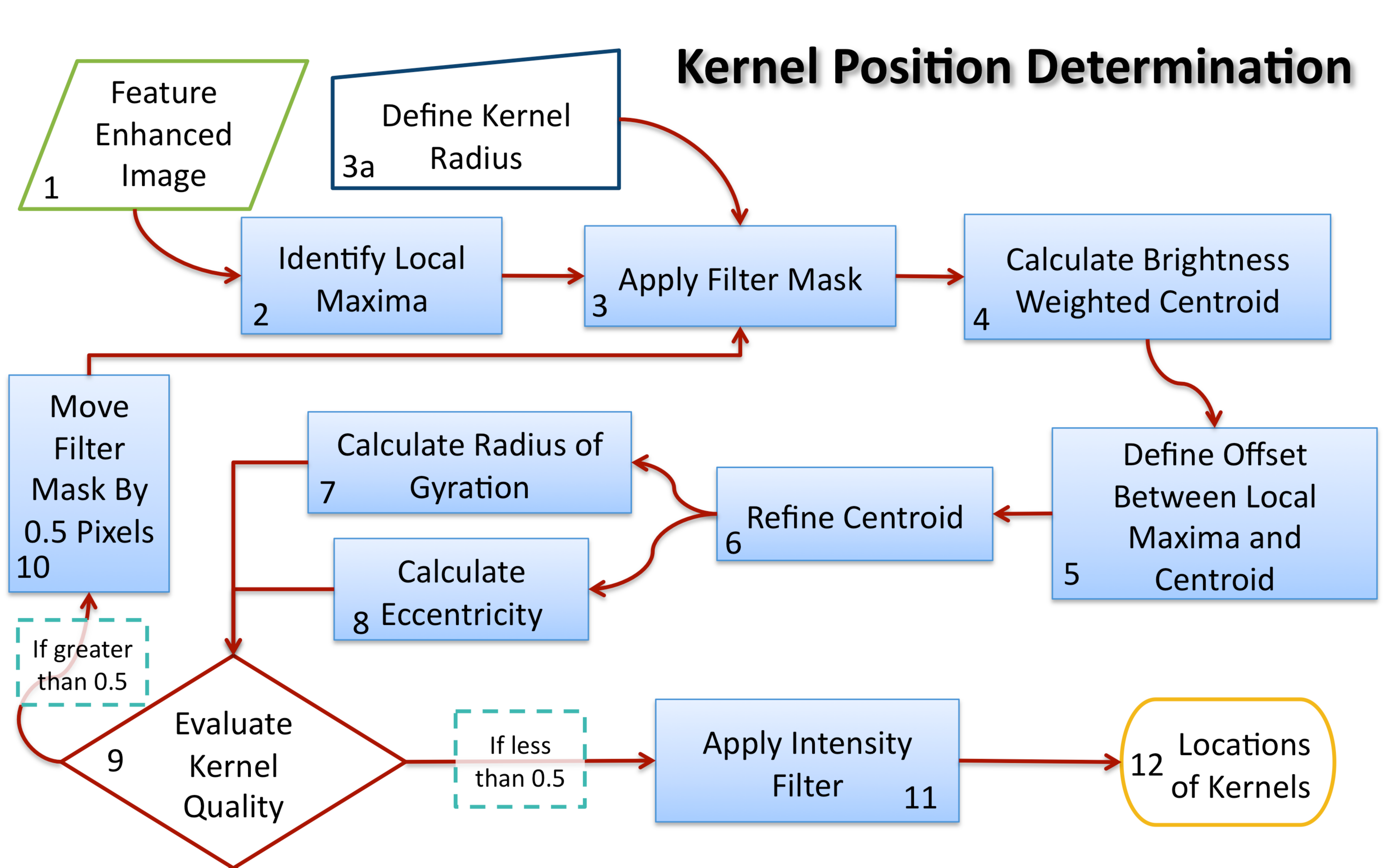}
              }
              \caption{A flowchart detailing the process in which flare and SCB kernels are identified.  The green parallelogram represents the database input. Blue trapezoids indicate a feature specific quality that must be determined and input. The red diamond is a process that has a conditional result and the blue shaded rectangles are individual routines. 
                      }
   \label{Position_flow}
   \end{figure}

Determining the location of kernels is an iterative process. The corresponding flowchart is shown in Figure~\ref{Position_flow}. The process begins by identifying local intensity maxima within the enhanced image, $\mathcal{I'}$,  as candidate kernels.  A pixel is considered a candidate centroid of a kernel if a brighter pixel does not exist within a radius of $\omega$ pixels of the candidate. In general, $\omega$ is the characteristic radius of the kernel the algorithm is looking for. We required an $\omega$ to maximize the number of detections along a flare ribbon as well as the number of frames in which a detection could be tracked. Without specific knowledge of what size a flare kernel or SCB kernel could be, $\omega$ was iterated until both of the conditions were met. To maximize the number of detection regions and minimize the number of untraceable kernels, both quantities were plotted versus mask radius for three flares used in the training set. A mask with an 8 pixel radius ($\omega=8$) was consistently found to be the minimum radius where the diagnostic curves crossed. 

Next the brightness-weighted centroid of pixels,$({\bf{x}}, {\bf{y}})$, for each candidate kernel is calculated with the local maxima having pixel coordinates $(\bar{x},\bar{y})$ and intensity $\mathcal{A}(\bar{x},\bar{y})$: 
\begin{equation}
 \left( \begin{array}{c}
{\bf{x}} \\
{\bf{y}} \end{array} \right)
=\frac{1}{m_0} \sum_{i^2+j^2 \le \omega^2}
\left( \begin{array}{c}
\bar{x} \\
\bar{y} \end{array} \right)
\mathcal{A}(\bar{x}+i, \bar{y}+j).
\end{equation}
$m_0$ is the integrated intensity of the candidate kernel:
\begin{equation}
\label{eq_tot_inten}
m_0= \sum_{i^2+j^2 \le \omega^2}\mathcal{A}(\bar{x}+i, \bar{y}+j)
\end{equation}
and $\omega$ is the radius of the mask which was defined above \cite{Crocker1996}. Improper background subtraction can bias the centroid toward the center of the mask and away from the kernel's brightness center by adding weight to pixels that are not part of the features being identified.  

An offset between the brightness weighted centroid and the local maxima is then defined by subtracting the position of the brightness centroid from the position of the local intensity maximum:
\begin{equation}
\label{offset}
 \left( \begin{array}{c}
\epsilon_x \\
\epsilon_y \end{array} \right)
= \left( \begin{array}{c}
{\bar{x}} \\
{\bar{y}} \end{array} \right) -
 \left( \begin{array}{c}
{\bf{x}} \\
{\bf{y}} \end{array} \right).
\end{equation}

Next, a refined centroid of the candidate kernel is calculated by: 
\begin{equation}
(x', y')=\left(\bar{x}+\frac{\epsilon_x}{2},\bar{y}+\frac{\epsilon_y}{2}\right).
\end{equation}
This refined centroid is located halfway between the local intensity maximum and the brightness-weighted centroid. 
 
To evaluate the quality of the candidate kernel, the discrepancy between the local maximum and centroid is examined. If either $|\epsilon_x|$ or $|\epsilon_y|$ (the offset as described by Equation~\ref{offset}) is greater than 0.5 pixels from the local maximum, the filtering mask is shifted by 0.5 pixels in the direction of $(\epsilon_x,\epsilon_y)$ and the new kernel is characterized. This iteration has the effect of guaranteeing that the local maximum and the centroid of the kernel are within half a pixel of each other. 

\begin{figure}    %%%%%%%%%%%%%%%%%% FIGURE 6 
   \centerline{\includegraphics[width=1.3\textwidth,clip=,angle=90]{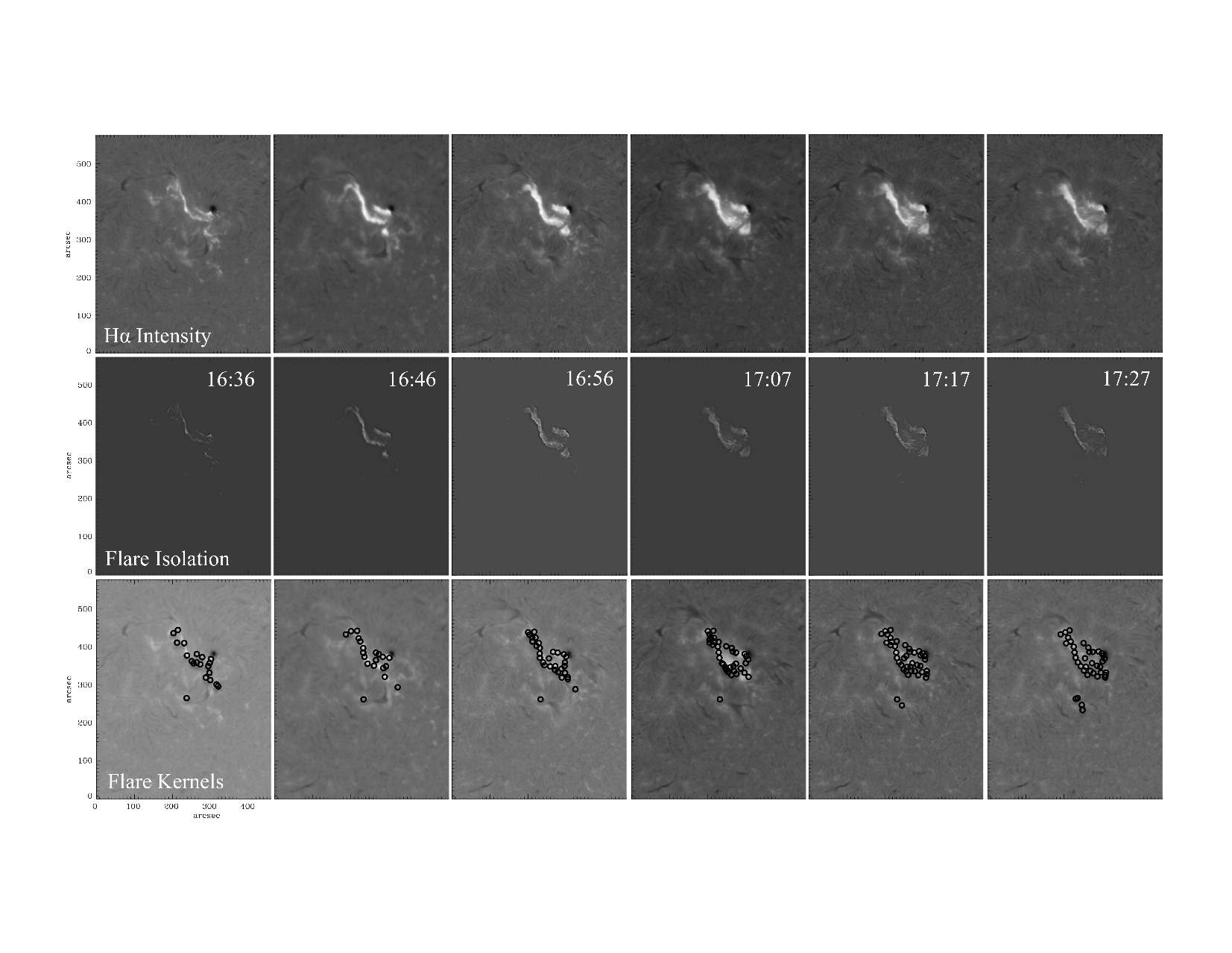}
              }
              \caption{A time series of images showing the evolution of the flare ribbons for the May 13, 2005 event. Each image shows the same flaring region ten minutes after the previous image. The flare peak occurs at 16:49 UT.  The top row of images show the H$\alpha$ line center intensity. The second row of images show the flare after the ribbon isolation and enhancement as described in Section~\ref{S-Flare_anal}. The third row of images show the detected flare kernels after the image analysis algorithm is applied. The centroid of each flare kernel is marked by a black circle. Notice how the number of kernels change as the flare ribbon evolves. A comparison of the first and third row of images show the coverage of detections along the flare ribbons. 
               }
   \label{Flare_kernels}
   \end{figure}

The final step in determining the kernel's position is to eliminate unwanted kernel detections or ``false kernels''. A threshold is applied to the integrated intensity of each candidate kernel. This intensity filter is intended to eliminate large, dim detections that may have a local maximum but are not part of the features in which we are interested. The threshold for this filter was determined empirically to eliminate marginal detections while retaining the bulk of the detections. Each detection now has position $(x'_i, y'_i)$ and an integrated intensity $m_{0,i}$ to characterize the kernel. The results of this detection process can be seen in the lower panels of Figure~\ref{Flare_kernels}. 

\subsubsection{General Trajectory Linking}
\label{S-trajectory} 
After the positions of all kernels within the series of images have been located, the next step is to associate kernels between frames to form trajectories. This presents a problem in determining which kernel in image $\mathcal{I'}_{n+1}$ is most likely to correspond to a given kernel in the preceding image, $\mathcal{I'}_{n}$.  It is complicated by the number of particles that are being tracked simultaneously. We simplify the problem by requiring that each kernel be identified with at most one kernel in the previous image. Since kernels are nominally indistinguishable from each other, a kernel pair's physical proximity between frames is a reasonable way to establish a correlation between the two. The following algorithm for trajectory linking is motivated by the dynamics of noninteracting Brownian particles; see \inlinecite{Crocker1996} for a more complete discussion. 

The probability that a single particle with Brownian motion will diffuse a distance $\delta$ in time $\tau$ is 
\begin{equation}
P(\delta|\tau)=\frac{1}{4\pi \mathcal{D} \tau} \exp \left(-\frac{\delta^2}{4\mathcal{D} \tau}\right)
\end{equation}
where $\mathcal{D}$ is the self-diffusion coefficient of each particle \cite{Crocker1996}. For a system of $N$ noninteracting particles, the diffusion probability distribution is the product of the distribution for each single particle:
\begin{equation}
P(\{\delta_i\}|\tau)=\left(\frac{1}{4\pi \mathcal{D} \tau}\right)^N \exp \left( -\sum_{i=1}^N \frac{\delta_i^2}{4\mathcal{D} \tau}\right).
\end{equation}
Thus the best assignment of trajectories is one that maximizes $P(\{\delta_i\}|\tau)$ or minimizes $ \sum{\delta_i^2}$ \cite{Crocker1996}. While merging or converging kernels result in a loss of tracking memory, we are able to visually discern any such occurrences by animating the trajectories. 

Calculating $P(\{\delta_i\}|\tau)$ for every possible combination of trajectories would need $O(N!)$ computations which is unfeasible and not necessary in this case. We know that a flare or SCB is going to stay fairly localized and not jump sporadically from one location to another. Thus we can define a characteristic diffusion length $L$ to be the maximum distance a kernel can travel between frames. This diffusion length is different between flares and SCBs (see Sections~\ref{S-Flare_anal} and~\ref{S-SCB_anal} for the specific differences).  With a small enough diffusion length, most kernels have only one possible association between two frames. 

If all kernels remained present throughout the series of images, then associating trajectories would be significantly simpler. Unfortunately this is not the case at all since we are interested in tracking the flare from its inception through decay. To account for a kernel disappearing and reappearing, ``missing'' links are assigned to a kernel that does not have a counterpart in the next frame. A diffusion distance of $\delta=L$ is assigned for the purposes of calculating $P(\{\delta_i\}|\tau)$. The last known location of a ``missing'' particle is retained for a predetermined number of frames. That number of frames is different for flares and SCBs (see Sections~\ref{S-Flare_anal} and~\ref{S-SCB_anal} for the specific differences). 

Linking detections into trajectories is only possible if the diffusion distance $\delta$ is significantly smaller than the typical spacing between kernels $a$. If $\delta$ is greater than $\sim \frac{a}{2}$ then trajectories are inextricably confused between images.  This property of confusion provides a upper bound for kernel motion which provides insights into physical properties (see Section~\ref{S-Results}).

Linking the initial kernel detections into trajectories provides another opportunity to filter out inconsistent or unsustained detections. A kernel lasting only one or two frames is a weak detection and can be filtered out. A temporal trajectory filter eliminates fleeting bright points that appear in a few frames but are not associated with feature being tracked. The specifics of this filter appears in Sections~\ref{S-Flare_anal} and~\ref{S-SCB_anal}. 

The last step in the kernel identification and tracking algorithm is to extract physical properties of the data.  To accomplish this extraction, the locations and parameters of the kernel trajectories are placed over the preprocessed dataset. Since the original H$\alpha$ images were modified to aid in the detection process in thresholding and feature enhancement, the tracked kernels are placed over the un-enhanced H$\alpha$ images to extract the intensity underneath the each kernel. Readers who are interested in applying these methods are welcome to contact the lead author.

\subsection{Flare Ribbon Extraction}
	\label{S-Flare_anal}
Within the general feature detection and tracking framework presented above and outlined in Figure~\ref{flow_chart}, there are several situations where the algorithm is customized to the features that it is trying to extract. This subsection will explain the specifics for extracting flare ribbons from the H$\alpha$ dataset. 

The first flare-specific customization is applying a threshold level of 1.35 to the preprocessed images where quiet Sun disk center intensity is set to unity. This eliminates any features that are approximately less than one standard deviation above background intensity.  This corresponds to keeping features with a 35\% increase in brightening over the background intensity and since all images are normalized to their background intensity, the threshold value holds true for any ISOON data set.

Next, the flares are isolated and enhanced. To isolate flares, a Laplacian operator is applied to the thresholded images. In this application, the Laplacian operator acts as an edge enhancement algorithm making smooth transitions on the image disappear while rapid changes in intensity stand out. The images operated upon are then added back into the original image,
\begin{equation}
\mathcal{I'}=\mathcal{I}+\nabla^2\mathcal{I},
\end{equation}
where $\mathcal{I}$ is the image after thresholding and the bandpass filter, and $\mathcal{I'}$ is the resultant image used for the detection of flare kernels (Figure~\ref{Flare_kernels}). By adding the Laplacian image back to the original, the edges and peaks in the flare ribbons are enhanced while the physical dimensions of the active regions stay the same. 

When linking flare kernels into trajectories, we define a characteristic diffusion length $L$ as the maximum distance a kernel can travel between frames. For flare kernels, $L=4$ pixels. This diffusion length was empirically found to be the maximum distance a flare kernel could travel to minimize the confusion of tags when associating kernels between images. The typical spacing between flare kernels, $a$, is approximately 8 pixels, thus $L \approx \frac{a}{2}$.  

Lastly, a filter is applied to remove unwanted trajectories by evaluating their temporal integrity. The previous 40 frames are considered when linking flare kernels into trajectories. This means that a kernel can go ``missing'' for 40 images and be associated together. Using the fact that flare ribbons are relatively long-lived phenomena, lasting more than an hour, a filter is applied to eliminate short-lived detections.  Any trajectories that last less than 20 frames are eliminated. This filter eliminates several off-ribbon detections which are associated with the eruption but do not characterize the evolution of the flare ribbons. 

\subsection{SCB Extraction}
	\label{S-SCB_anal}
SCB detections are completely eliminated in the flare detection and tracking process.  A separate process is required to extract SCB kernels and characterize this different type of brightening that occurs during a flare eruption. This subsection details the customizations to the detection algorithm needed to extract SCBs. 
 
To begin to isolate the SCBs from the original background normalized images, a threshold of 1.2 is taken.  This threshold eliminates all features except where intensities greater than 20\% above background level.  

\begin{figure}    %%%%%%%%%%%%%%%%%% FIGURE 7 
   \centerline{\includegraphics[width=1.3\textwidth,clip=,angle=90]{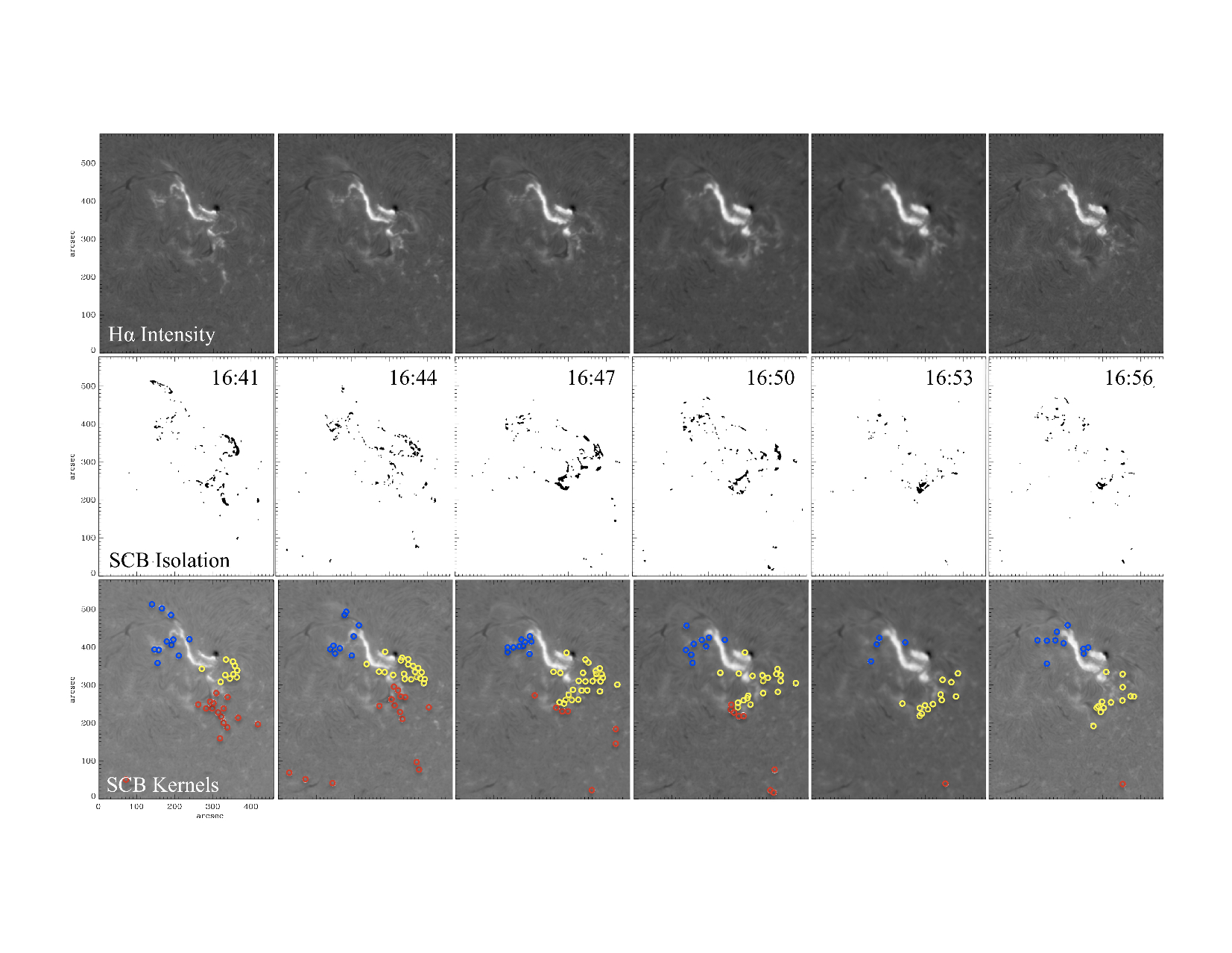}
              }
              \caption{A time series of images showing the evolution of the SCBs for the May 13, 2005 event. Each image shows the same flaring region three minutes after the previous image. The flare peak occurs at 16:49 UT.  The top row of images show the H$\alpha$ line center intensity (same as the top row of Figure~\ref{Flare_kernels}). The second row of images show a binary image of the same flaring region after applying the SCB isolation as described in Section~\ref{S-SCB_anal}. The third row of images show the detected SCB kernels after the image analysis algorithm is applied. Identified SCB's were visually separated into three groups to demonstrate how they appear to move when watching an animation of the flare. The red group propagates quickly while the blue and yellow groups develop more slowly.  Notice how the number of kernels peak before the flare peak and rapidly declines as compared to the flare kernels.  
               }
   \label{SCB_kernels}
   \end{figure}

To isolate the SCB features in a series of H$\alpha$ images, a running difference image series is created.  In general a running difference image series eliminates long-lived features while enhancing rapidly changing intensities. In this case, the current image in the series is a binary mask of all the pixels above the 1.20 threshold.  The binary image is necessary to force all brightening above the threshold to a uniform value and includes both SCBs as well as flare ribbons. Next, a median image is defined by taking a median of the previous 11 frames. The running difference image is calculated by subtracting the median image from the current binary image, 
\begin{equation}
\mathcal{I'}_n=\mathcal{I}_n-\widetilde{\mathcal{I}}_{n-11:n-1},
\end{equation}
where $\mathcal{I}$ is the binary image,  $\widetilde{\mathcal{I}}$ is the median image, and $\mathcal{I'}$ is the running difference image.  The 11 minute duration of the median image was experimentally determined to best maximize the number of SCBs detected over the three training flares examined. The results of this process expose the locations of the SCBs while subtracting off the longer-lived flare ribbons (Figure~\ref{SCB_kernels}).

The isolated SCBs are then enhanced through a morphological transform.  The transform performs an erosion operation followed by a dilation operation with a 3$\times$3 pixel plus-shaped structuring element. This transform has the effect of removing small regions of noise in the running difference image while preserving the size and shape of the features in the image.  

Using the binary running difference images, the SCB's local maxima are identified as outlined in Section~\ref{kernel_position}. However since the images being analyzed are binary images, the local maximum is coincident with the centroid of the bright points within the mask which significantly decreases processing time. The kernels are then run through the trajectory linking algorithm with a diffusion length $L=10$. This diffusion length allows the centroid of the SCB kernel to move no more than 10 pixels between frames. 

When correlating SCB kernels between frames into trajectories, only the previous image is searched. A temporal filter is also applied to the trajectories requiring at least two frames. What remains are kernels that last at least two minutes and do not ``go missing'' at anytime. Lastly, a spatial filter is applied to eliminate any SCB kernels that are identified as within the ``flaring region.'' This region is defined by co-adding 80 of the thresholded images used in the identification of flare kernels centered around the peak of the flare intensity. All non-zero pixels are flagged as the ``flaring region''. The spatial filter guarantees that any SCB kernels that are detected are not associated with the flare ribbons. The resulting detections are shown in Figure~\ref{SCB_kernels}.

\section{Physical Results of Algorithm Application}
	\label{S-Results}
The kernel tracking algorithm described in the previous section is applied to three different cases to train and refine the parameters. Table~\ref{T-events} outlines the basic characteristics of these datasets. From this initial application of the tracking algorithm, several interesting results about flare and SCB kernels presented themselves. Section~\ref{S-Flare_results} discusses the derived physical properties of flare kernels and Section~\ref{S-SCB_results} characterizes the physical parameters of SCB kernels. Table~\ref{T-results} provides a summary of the results described in this section. 

\subsection{Properties of Flare Kernels}
\label{S-Flare_results}
We define a characteristic diffusion length $L$ for flare kernels to maximize the length of trajectories formed while minimizing confusion between kernel tags (see Section~\ref{S-trajectory}). This parameter $L$ corresponds to the maximum distance a kernel can travel between frames. A parameter of $L =4$ pixels is found to be the best diffusion length for flare kernels.  Thus, the maximum physical distance that a flare kernel can travel between frames translates to 3,200 km. Since $L \sim \frac{a}{2}$, where $a$ is the characteristic distance between kernels, the diameter of the smallest resolvable kernel along the flare ribbon is approximately 6,400 km with the ISOON dataset. 

After flare kernels are associated between frames into trajectories, the temporal integrity of flare kernels can be exposed. Flare kernels are temporally fairly robust; lasting on average $\sim70$ minutes each. Generally, large flares take a few hours to evolve from the impulsive phase, through peak intensity, and return to pre-flare brightness. Also, the number of detectable kernels declines as the flare's intensity decays from its peak implying that there are fewer resolvable components in the flare ribbons as the flare evolves.  This change in detectable kernels means that the majority of kernels cannot be tracked from pre-flare to post-flare suggesting a dynamic substructure to the flare ribbons when bright points appear and disappear as the flare erupts. 

\begin{figure}    %%%%%%%%%%%%%%%%%% FIGURE 8 
   \centerline{\includegraphics[width=1.2\textwidth,clip=,angle=90]{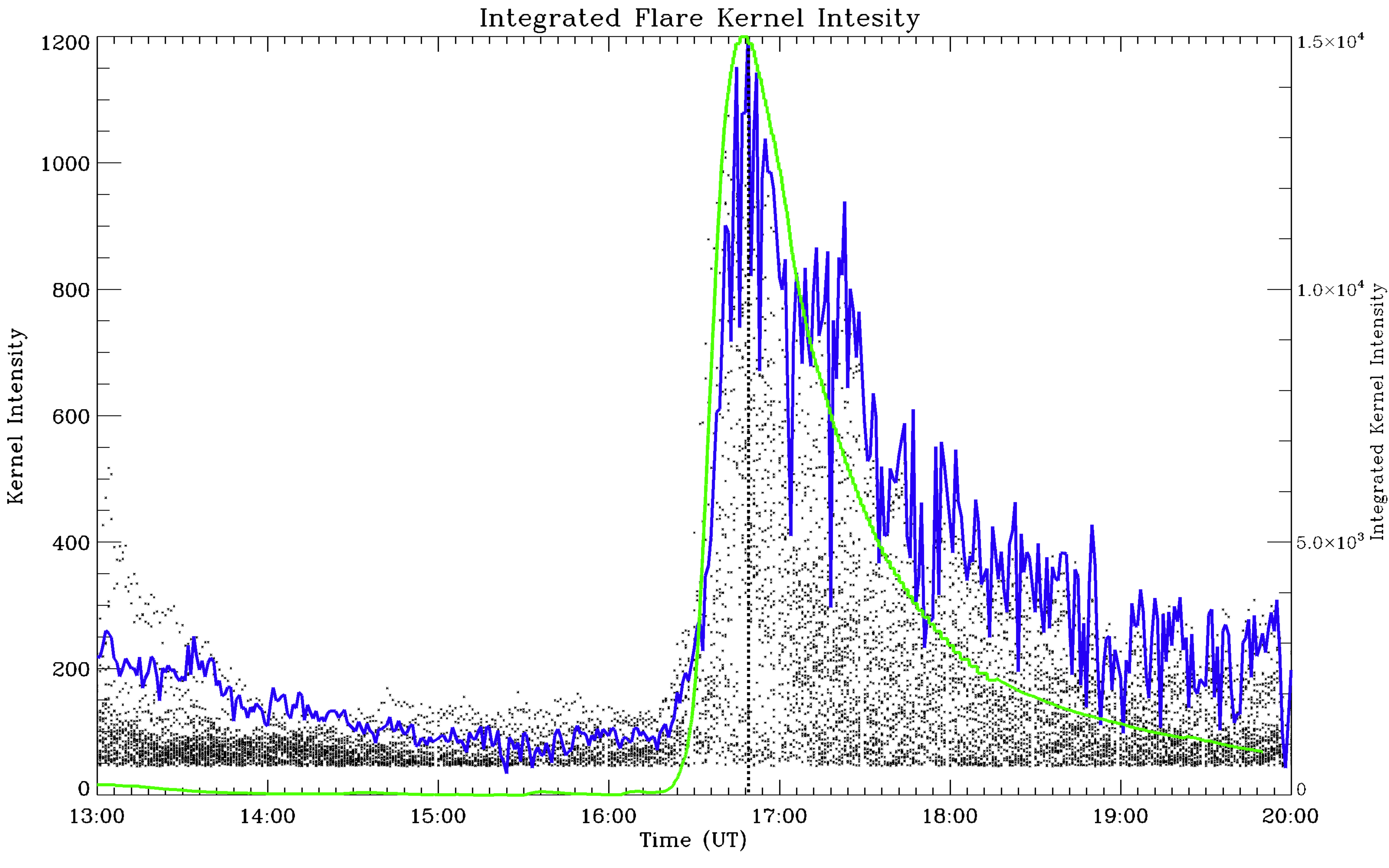}
              }
              \caption{A plot showing the May 13, 2005 measured intensities of each flare kernel plotted as points as a function of time. The solid blue line is the integrated kernel intensities at each time step. Overlaid for comparison in green is the normalized GOES $1.0-8.0$ \AA intensity curve. Here again, the vertical dashed line indicates the peak of the flare curve as shown in Figure~\ref{0513_boxes}{\bf C}. The integrated flare kernels reproduces the shape of the flare curve determined by other methods. This indicates that the flare is well characterized by the flare kernels when taken in aggregate. 
               }
   \label{Integrated_flare}
   \end{figure}

Because flare kernels do not last for the entirety of the flare, individual kernels are not a good indication of overall flare behavior. However when taken in aggregate, flare kernels reproduce the overall topology of the flare. Figure~\ref{Integrated_flare} shows the intensities of all of the tracked flare kernels' plotted as points as a function of time.  Integrating these intensities over each time step yields an aggregate intensity curve that reproduces the flare intensity curve shown in Figure~\ref{0513_boxes}~{\bf C}. 

\subsection{Properties of SCB Kernels}
\label{S-SCB_results}
\begin{figure}    %%%%%%%%%%%%%%%%%% FIGURE 9
   \centerline{\includegraphics[width=1.2\textwidth,clip=,angle=90]{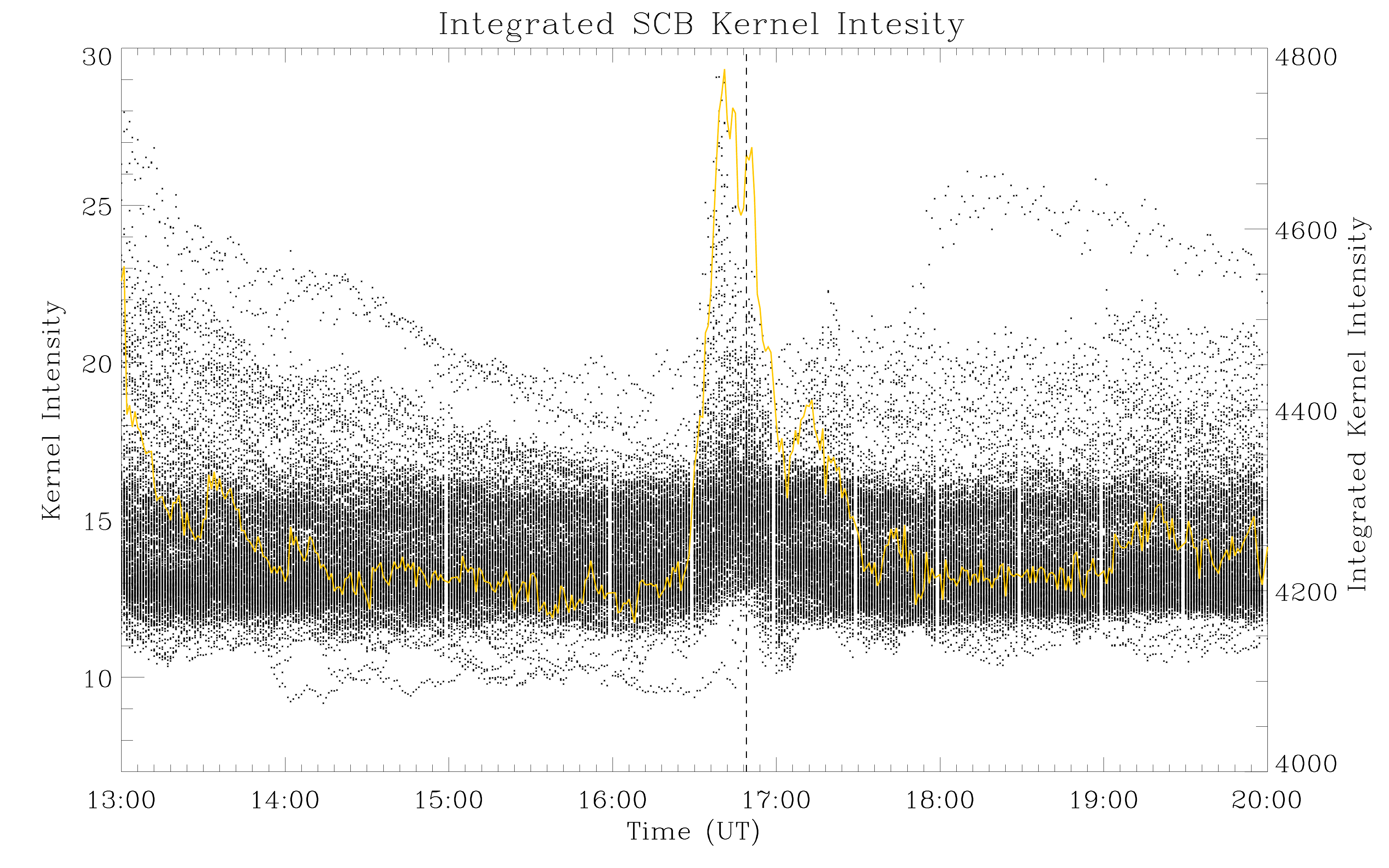}
              }
              \caption{A plot showing the May 13, 2005  measured intensities of each SCB curve plotted as points as a function of time. The solid yellow line is the integrated kernel intensities at each time step. Here again, the vertical dashed line indicates the peak of the flare curve as shown in Figure~\ref{0513_boxes}{\bf C}. Notice that the intensity of the integrated SCB curve returns to the continuum level much faster than the integrated flare kernel curve as shown in Figure~\ref{Integrated_flare}. This indicates that the SCB kernels are measuring something distinctly different than the flare kernels. Also notice how the SCB intensities peak $\sim$ 30 minutes before the flare intensity peaks.
               }
   \label{Integrated_SCB}
   \end{figure}

Sequential chromospheric brightenings, although related to the erupting flare ribbons, are distinctly different than the flare kernels. Figure~\ref{Integrated_SCB} plots the intensity of the entire population of SCBs versus time. It shows a relatively small time window when SCB occur before returning to pre-flare intensities. SCBs begin brightening about 30 minutes before flare peak and return to background intensity about 20 minutes after. In contrast, the flare intensity curve remains above pre-flare levels for several hours. 

Another result that is well demonstrated when plotting SCB kernel intensities in aggregate, Figure~\ref{Integrated_SCB}, is SCBs never have a peak intensity greater than their host flare. An SCB has a peak intensity 1.2 -- 2.5 times above the background intensity level. In contrast, flares can brighten an order of magnitude or more above the pre-flare brightness. 

Individual SCBs are also much more fleeting than the flare ribbons. They increase to their peak brightness and return to background intensities in a much shorter timescale than the flare ribbons are observed to do.  When isolated, these ephemeral phenomena have intensity curves that are impulsive with a sharp peak and then a return to background intensity in the span of about 12 minutes. 

When the tracks of SCB kernels are examined, they do not show any progressive motion. The centroid of an SCB kernel randomly walks around within about 6 pixels of its starting location for the duration of the trajectory. Although SCB's sequential nature of point brightening gives the appearance of a progressive traveling disturbance, the plasma beneath each brightening does not follow the disturbance and remains in the same location.  This result confirms the findings of \inlinecite{Bala2005}. The third row of images in Figure~\ref{SCB_kernels} are manually color-coded to reflect this traveling disturbance that is seen in an animation of the detections. The red group of detections matures quickly and dissipates while the yellow and blue groups take more time to develop. This suggests that there may be more than one population of SCBs. 

\begin{table}
\caption{ 
A summary of results for flare and SCB kernels.  
}
\label{T-results}

\begin{tabular}{lcccc}     % define the column alignment
  \hline                   % horizontal line
Kernel & Minimum Diameter & Maximum Intensity & Average Lifetime & Motion \\
  & (km) & (above background) & (minutes) & \\
  \hline
Flare & 6,400 & 10 & $\sim$ 70 & Directional \\
SCB & 1,600 & 1.5 & $\sim$ 12 & Random \\
  \hline
\end{tabular}
\end{table}

\section{Discussion}
	\label{S-Discuss}
The method and results presented here suggest tracking flare kernels through the evolution of the erupting flare can characterize the evolving active region.  We also demonstrate that a sum of the components of the flare ribbon reproduces the total intensity curve of the flare. Although it is not possible to say that any given kernel is tracking one specific flare loop, the flare kernels dissect the flare into its smallest visibly resolvable components in the ISOON H$\alpha$ dataset. Since the number of detectable kernels declines as the flare's intensity decays, the overall intensity of the flare is related to the number of ribbon components discernible at any given time.	

The tracking algorithm also confirms the results of \inlinecite{Bala2005} that sequential chromospheric brightenings are physically different from their associated flare ribbons. The tracked SCB kernels do not physically progress in one direction although the appearance of SCBs are sequential and propagate out from the flare ribbon. SCB kernels demonstrate an impulsive brightening without an exponential decay.  Both of these physical characteristics of SCBs are distinctly different than the associated flare ribbons which do appear to physically move and have sustained intensity brightening lasting an hour or more. These differences imply a distinctly different physical mechanism is causing the sequential brightening as opposed to the flare ribbon brightening which confirms the assertion made by \inlinecite{Pevtsov2007}.

One major advantage of tracking flares and associated brightening with this method is that the locations and tracks of the flare and SCB kernels can be overlaid onto any dataset that is co-aligned with the original H$\alpha$ images.  A more thorough and systematic examination of flares and SCBs is now possible allowing the user to develop statistics on any number of flares. Two ribbon flares without previously identified SCB's can also be analyzed with this automated routine to further investigate if SCB's are limited to moderate sized flares with associated CME's. To obtain a copy of this code, please contact the corresponding author with your request.

\begin{acks}
 The authors thank: (1) USAF/AFRL Space Scholar Program, (2) NSO/AURA for the use of their Sunspot, NM facilities, (3) AFRL/RVBXS, (4) Crocker and Weeks for making their algorithm available online, and (5) NMSU. This project was supported by grants from the Air Force Office of Scientific Research and New Mexico Space Grant Consortium. 
\end{acks}

     % format of references provided by the journal (.bst)
\bibliographystyle{spr-mp-sola}
%\bibliographystyle{spr-mp-sola-cnd} %% Alternative style: no title,
                                      % no concluding page. 

     % name your Bibtex file containing your references (.bib)
\bibliography{SCB_Methods_Bibliography}

%==========FIGURES=================
%------------------------------

%------------------------------
   
 %------------------------------  

%------------------------------  

%------------------------------

%------------------------------

   %------------------------------
   
  %------------------------------

\end{article} 

\end{document}